\documentclass[lettersize,journal]{IEEEtran}
\usepackage{amsmath,amsfonts}
\usepackage{array}
\usepackage{textcomp}
\usepackage{stfloats}
\usepackage{url}
\usepackage{verbatim}
\usepackage{graphicx}
\usepackage{caption}
\usepackage{subcaption}
\usepackage{cite}
\usepackage{makecell}
\usepackage{multirow}
\usepackage{booktabs} 
\usepackage{longtable}
\usepackage{stfloats}  

\usepackage{algorithmic}
\usepackage{algorithm}
\usepackage{amssymb}
\usepackage{color}

\begin{document}
	\title{Modulation Consistency-based Contrastive Learning for Self-Supervised Automatic Modulation Classification}
	\author{
		Chenxu Wang,  
		Shuang Wang, \IEEEmembership{Senior Member,~IEEE,} 			
		Lirong Han, 
		Xinyu Hu,
		Hanlin Mo, 	
		Hantong Xing,		\\
		Sotiris A. Tegos, \IEEEmembership{Senior Member,~IEEE,} 
		and Licheng Jiao, \IEEEmembership{Life Fellow,~IEEE}
		
		\thanks{This work was supported in part by the National Natural Science Foundation of China under Grant 62271377 and Grant 62201407; in part by the National Key Research and Development Program of China under Grant 2021ZD0110400 and Grant 2021ZD0110404; in part by the Key Research and Development Program of Shannxi Program under Grant 2021ZDLGY01-06, Grant 2022ZDLGY01-12, Grant 2023YBGY244, Grant 2023QCYLL28, Grant 2024GX-ZDCYL-02-08, and Grant 2024GX-ZDCYL-02-17.
			(\textit{Corresponding author: Shuang Wang})
		}
		
		\thanks{
			Shuang Wang, 
			Chenxu Wang, 
			Lirong Han, 
			Xinyu Hu, and Licheng Jiao are with the School of Artificial Intelligence, Xidian University, Xi’an 710071, China. (email: wcxlocal@163.com; shwang@mail.xidian.edu.cn; hanlirong@xidian.edu.cn; huxinyu0410@stu.xidian.edu.cn; lchjiao@mail.xi
			dian.edu.cn).					
			Hanlin Mo is with the Unmanned System Research Institute, Northwestern Polytechnical University, Xi’an 710072, China (e-mail: mohanlin@nwpu.edu.cn).
			Hantong Xing is with the School of Electronics And Information, Northwestern Polytechnical University, Xi’an 710072, China (e-mail: xinghantong@126.com).			
			Sotiris A. Tegos is with the Department of Electrical and Computer Engineering, Aristotle University of Thessaloniki, 54124 Thessaloniki, Greece (e-mail: tegosoti@auth.gr).
		
		}
		
	}

	\markboth{Journal of \LaTeX\ Class Files,~Vol.~14, No.~8, July~2025}%
	{Shell \MakeLowercase{\textit{et al.}}: Modulation Consistency-based Contrastive Learning for Self-Supervised Automatic Modulation Classification}

	\maketitle

	\begin{abstract}
		Deep learning-based AMC methods have achieved remarkable performance, but their practical deployment remains constrained by the high cost of labeled data.
		Although self-supervised learning (SSL) reduces the reliance on labels, existing SSL-based AMC methods often rely on task-agnostic pretext objectives misaligned with modulation classification, leading to representations entangled with nuisance factors such as symbol, channel, and noise.
		In this paper, we identify \textit{intra-instance modulation consistency} as a task-aware structural prior, whereby different temporal segments of the same signal may differ in waveform while preserving the same modulation type, thus providing a principled cue for task-aligned self-supervision.
		Based on this prior, we propose Mod-CL, a Modulation consistency-based Contrastive Learning framework that constructs positive pairs from different temporal segments of the same signal instance, to encourage the model to learn shared modulation information while suppressing nuisance variations.
		We further develop a contrastive objective tailored to Mod-CL, which jointly exploits temporal segmentation and data augmentation to pull together views sharing the same modulation semantics while avoiding supervisory conflicts within each signal instance.
		Extensive experiments on RadioML datasets show that Mod-CL consistently outperforms strong baselines, especially in low-label regimes, achieving substantial improvements in linear probing accuracy.
		
	\end{abstract}
	

\section{Introduction}\label{sec:intro}

\IEEEPARstart{W}{ith} the rapid development of wireless communication technologies and the widespread deployment of Internet of Things (IoT) devices, achieving efficient and secure communications among intelligent devices has become increasingly challenging\cite{11002534,zheng2025recent}. 
Automatic modulation classification (AMC), which aims to identify the modulation type of a received signal by analyzing its characteristics, plays a crucial role in ensuring communication reliability\cite{10753459, 11153513}, improving spectral efficiency\cite{rajendran2018deep, zhao2024ultra}, and enhancing signal awareness\cite{hao2023contrastive,10422963}.

Existing AMC approaches can be broadly categorized into traditional methods and deep learning (DL)-based methods\cite{zheng2025recent}. 
Traditional approaches rely on handcrafted signal features~\cite{10502319,10499712} or statistical models~\cite{dulek2017online,zhu2018likelihood}, which may achieve satisfactory performance under specific conditions but often exhibit limited generalization in dynamic wireless environments. 
Recently, DL-based AMC methods have demonstrated superior performance by automatically learning discriminative feature representations through end-to-end training, thereby avoiding the need for manual feature engineering and exhibiting improved robustness under complex channel conditions\cite{9058656,9576081}. 	
Despite the progress achieved by DL-based AMC methods, their reliance on large-scale, accurately labeled datasets remains a key limitation for practical deployment. 	
In realistic wireless communication scenarios, obtaining reliable modulation labels is often costly and, in some cases, even infeasible~\cite{10562208,kong2023transformer}. 
This challenge motivates the exploration of representation learning paradigms that can extract discriminative and robust features from unlabeled data, thereby mitigating the reliance on extensive manual annotations.

Self-supervised learning (SSL) has emerged as a promising paradigm for AMC by leveraging unlabeled signals through pretext tasks, especially in realistic and non-cooperative scenarios where reliable labels are scarce~\cite{chen2020simple,he2022masked}. 
Existing SSL-based AMC methods typically adopt \textit{task-agnostic} pretext objectives to learn representations, such as contrasting augmented views~\cite{liu2021self,kong2023transformer} or reconstructing masked observations~\cite{shi2023gaf,xu2025predicting}.
While effective in capturing signal global structure, these objectives do not explicitly enforce alignment with the modulation classification task.
However, a received signal is influenced by multiple latent generative factors in practical wireless environments, including the modulation scheme, symbol sequence, channel effects, and noise~\cite{8054694,RML22}. Among these, the modulation scheme constitutes the primary discriminative factor for AMC, 
whereas the others can be regarded as nuisance variations with respect to the modulation classification objective under typical modeling assumptions.
When such factor distinctions are not explicitly considered, SSL models may encode both modulation-relevant and irrelevant information, leading to entangled representations. 
As a result, the learned invariances are not necessarily aligned with modulation semantics, which may limit downstream AMC performance.

\begin{figure}
	\centering
	\includegraphics[width=0.85\linewidth]{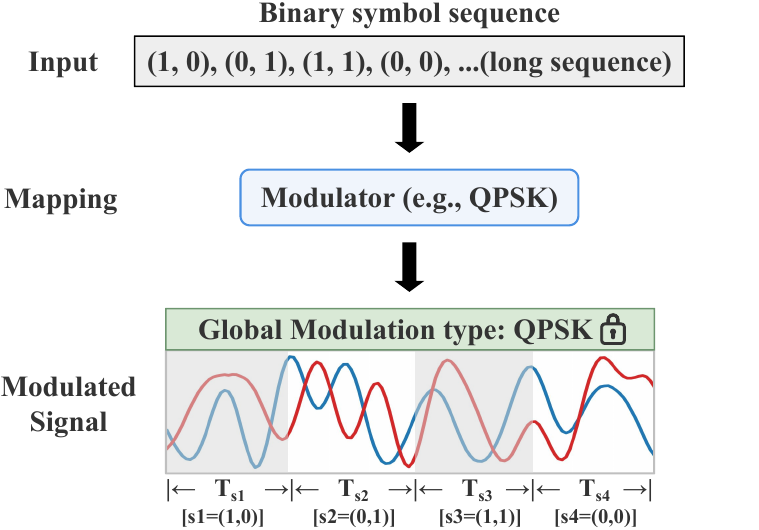}
	\caption{
		Schematic illustration of intra-instance modulation consistency in the modulation process.
		A bit sequence is mapped by a fixed modulator, e.g., QPSK, to generate an IQ signal whose temporal portions correspond to different symbol realizations and thus exhibit distinct local waveform patterns, while sharing the same global modulation type.
	}
	
	\label{fig:motivation}
\end{figure}

This limitation highlights the need for a task-aware inductive bias that encourages modulation-aligned representation learning.
From the perspective of signal generation, each signal instance is associated with a fixed modulation scheme, which preserves its modulation identity within the instance~\cite{haykin1988digital,o2016radio}.
As illustrated in Fig.~\ref{fig:motivation}, the temporal portions $T_{s1}$--$T_{s4}$ correspond to different underlying symbol realizations, such as $(1,0)$, $(0,1)$, $(1,1)$, and $(0,0)$, and therefore exhibit distinct local waveform patterns.
However, these portions are generated by the same modulator and share the same global modulation type.
We term this phenomenon \textit{intra-instance modulation consistency} and view it as an inherent structural prior of modulated signals.
This prior provides a natural self-supervisory cue for contrastive learning by enabling positive-pair construction from segments that share modulation semantics while differing in local symbol-induced waveform patterns.
Since contrastive learning encourages representations to preserve information shared across positive pairs, such a pairing strategy better aligns the pretext objective with the downstream modulation classification task.
As a result, the learned representations are encouraged to emphasize modulation-related structures while reducing reliance on segment-specific symbol-induced waveform details.

Building upon this prior, we propose Mod-CL, a Modulation consistency-based Contrastive Learning framework for AMC.
Mod-CL exploits the intra-instance modulation consistency illustrated in Fig.~\ref{fig:motivation} by constructing positive pairs from different temporal segments of the same signal instance.
Such segments share modulation semantics while generally differing in local symbol-induced waveform patterns, thereby discouraging reliance on symbol-sequence-specific waveform details and encouraging the encoder to capture modulation-related structures shared across segments.
To optimize these structured positives, we propose a contrastive objective that decomposes the positive-pair relations into three complementary groups: segment consistency, augmentation consistency, and joint consistency. The segment consistency term aligns different temporal segments to enforce invariance to symbol-level variations; the augmentation consistency term promotes robustness to channel-style distortions; and the joint consistency term regularizes cross-view cross-segment pairs, enabling synergistic learning across both variations. 
This structured design explicitly encodes the intra-instance modulation prior, encouraging the encoder to focus on modulation-relevant features while being invariant to symbol-induced waveform variations.
We further provide a mechanism-level analysis to clarify the inductive bias of Mod-CL.
By modeling the signal generation process and analyzing the shared information between segment-level positive pairs, we find that contrastive learning weakens local symbol realizations that are not shared across segments, while biasing the learned representations toward stable cross-segment structures more aligned with modulation classification.

Extensive experiments on the RadioML datasets~\cite{o2016convolutional} demonstrate that Mod-CL consistently outperforms strong generic self-supervised and AMC-oriented baselines, especially under label-scarce settings and low-to-medium SNR conditions.
Beyond performance gains, semantic corruption and ablation analyses verify that the proposed modulation-aligned design effectively captures essential signal semantics while suppressing nuisance variations

The main contributions of this paper are summarized as follows:

\begin{itemize}
	
	\item[$\bullet$]
	A task-aware structural prior, termed \emph{intra-instance modulation consistency}, is identified for AMC, where different temporal segments of the same signal share modulation semantics while differing in local symbol realizations.	
	This prior provides a principled basis for label-free construction of modulation-consistent positive pairs.

	\item Mod-CL is proposed to exploit the identified prior by constructing positive pairs from different temporal segments of the same signal	instance. This design encourages the encoder to capture	modulation-related structures shared across segments while reducing its reliance on symbol-induced waveform details.

	\item A structured contrastive loss is developed to decompose the positive-pair relations induced by temporal segmentation and data	augmentation. By explicitly encoding the identified prior, this loss encourages modulation-invariant representations. An information-flow	analysis is further provided to explain the modulation-aligned inductive bias of the proposed framework.
	
	\item[$\bullet$]
	Extensive experiments on two public AMC benchmarks demonstrate the effectiveness of Mod-CL, with consistent gains over strong baselines, particularly under label-scarce settings. 
	Ablations validate the effectiveness of each component and the impact of hyperparameters.
	
\end{itemize}

	The remainder of this paper is organized as follows. 
	Section~II reviews related work on DL-based AMC and SSL for modulation classification. 
	Section~III formulates the signal model and problem setting, and presents the proposed Mod-CL, together with a mechanism-level analysis of its inductive bias. 
	Section~IV describes the experimental setup and benchmarks on public AMC datasets, followed by comprehensive results, ablations, and discussions. 
	Finally, Section~V concludes the paper and outlines directions for future work.

	\section{Related Work}

	\subsection{Automatic Modulation Recognition}	
	Automatic modulation classification (AMC) is a long-standing research problem in wireless communications, aiming to identify the modulation type of a received signal under various channel conditions.
	Early AMC approaches mainly relied on handcrafted signal features, such as higher-order cumulants\cite{orlic2009automatic,huang2016automatic}, cyclostationary statistics\cite{7036609,camara2019automatic}, and likelihood-based decision rules\cite{dulek2017online,zhu2018likelihood}, combined with classical statistical classifiers.
	These methods are often interpretable and can perform well under controlled settings, but their performance tends to degrade in practical scenarios with low signal-to-noise ratios, channel fading, and other non-ideal impairments.
	With the rapid development of DL, data-driven AMC methods have gained increasing attention.
	Convolutional neural networks\cite{10475703,chen2021signet}, recurrent neural networks\cite{rajendran2018deep,ke2021real}, and more recently transformer-based architectures\cite{hamidi2021mcformer,xu2025amc} have been employed to directly learn discriminative representations from raw IQ samples or time--frequency representations.
	These approaches generally achieve superior performance and robustness compared to traditional feature-based methods.
	However, their effectiveness typically relies on large-scale labeled datasets and extensive supervision, which are costly to acquire and often unavailable in realistic or non-cooperative wireless environments.

\subsection{Self-Supervised Learning for AMC}

To alleviate the reliance on labeled data, SSL has been actively explored for AMC by exploiting intrinsic structures of unlabeled signals.
Through appropriately designed pretext tasks, SSL aims to learn discriminative representations for modulation classification under limited supervision.
Existing SSL-based AMC approaches can be broadly categorized into reconstruction-based and contrastive learning-based methods.

Reconstruction-based methods guide representation learning by designing signal reconstruction pretext tasks.
Xu et al. employed frequency-domain representations as reconstruction targets, enabling models to learn implicit signal characteristics through cross-domain mapping between time and frequency domains\cite{xu2025predicting}.
Shi et al. adopted Gramian Angular Field representations to transform time-series signals into images and constructed a masked autoencoder for self-supervised reconstruction pretraining on unlabeled data\cite{shi2023gaf}.
These approaches demonstrate that reconstruction objectives can capture temporal correlation structures and improve representation quality for AMC to some extent.

Contrastive learning-based methods learn representations by pulling positive pairs closer while pushing negative pairs apart in the embedding space.
Liu et al. introduced rotation-based data augmentation as an early attempt to apply contrastive learning to AMC\cite{liu2021self}.
Kong et al. systematically investigated various data augmentation strategies and proposed a Transformer-based self-supervised framework with time-warp augmentations\cite{kong2023transformer}.
Li et al. further designed both intra-domain and inter-domain augmentations and employed a multi-representation domain-attentive contrastive learning scheme to enhance cross-domain feature extraction\cite{li2025multi}.

Overall, most existing SSL approaches for AMC are built upon contrastive learning with augmentation-defined positive pairs.
While effective, the invariances induced by such positive pairs are often implicit and may still retain nuisance factors such as symbol realizations, noise, and channel distortions.
In contrast, the proposed method defines positive pairs based on the signal generation mechanism itself, leveraging intra-instance modulation consistency to explicitly encourage modulation-aware invariance and suppress symbol-induced variations.

\section{Proposed Method}

\subsection{Signal Model and Problem Definition}
\label{sec:signal_model}

In a practical communication link, the received waveform is jointly influenced by the modulation scheme, symbol realizations, channel impairments, and noise.
We adopt a generic baseband observation model:
\begin{equation}
	\mathbf{x}_n = \mathcal{H}_n\!\left(\mathbf{s}_n(M_n,\mathbf{b}_n)\right) + \mathbf{w}_n,
	\label{eq:rx_model_generic}
\end{equation}
where $\mathbf{x}_n \in \mathbb{R}^{2\times T}$ denotes the $n$-th received IQ instance of length $T$,
$\mathbf{s}_n(M_n,\mathbf{b}_n)$ is the transmitted baseband signal generated under modulation scheme $M_n$ with a specific symbol realization $\mathbf{b}_n$,
$\mathcal{H}_n(\cdot)$ denotes an effective channel/distortion operator that may include fading and RF impairments,
and $\mathbf{w}_n$ represents additive noise.
From the signal generation perspective, the modulation scheme $M_n$ determines the mapping rule from symbols to baseband waveforms, while the symbol realization $\mathbf{b}_n$ evolves over time under this fixed mapping.
As a result, within a single received instance (i.e., a short observation window commonly considered in AMC), the modulation scheme $M_n$ can be reasonably regarded as fixed.
Meanwhile, the time-varying symbol realizations embedded in $\mathbf{s}_n(\cdot)$, together with the randomness of additive noise and channel effects, lead to noticeable waveform variations across different temporal segments of the same instance.

DL-based AMC can be formulated as a classification problem, where an encoder $f_{\theta}(\cdot)$ and a classifier $c_{\omega}(\cdot)$ are learned to predict the modulation label:
\begin{equation}
	\hat{y}_n = c_{\omega}\!\left(f_{\theta}(\mathbf{x}_n)\right), \qquad y_n=\pi(M_n)\in \{1,\dots,K\},
	\label{eq:amc_dl}
\end{equation}
where $K$ denotes the number of modulation classes and $\pi(\cdot)$ maps a modulation scheme to its class index.

In scenarios where labeled data are scarce or unavailable, SSL first pretrains the encoder $f_{\theta}(\cdot)$ on unlabeled signals via a pretext task $\mathcal{T}$, and then evaluates or adapts it for downstream AMC through linear probing and/or fine-tuning with limited labeled data.

\begin{figure*}[t]
	\centering
	\includegraphics[width=1\linewidth]{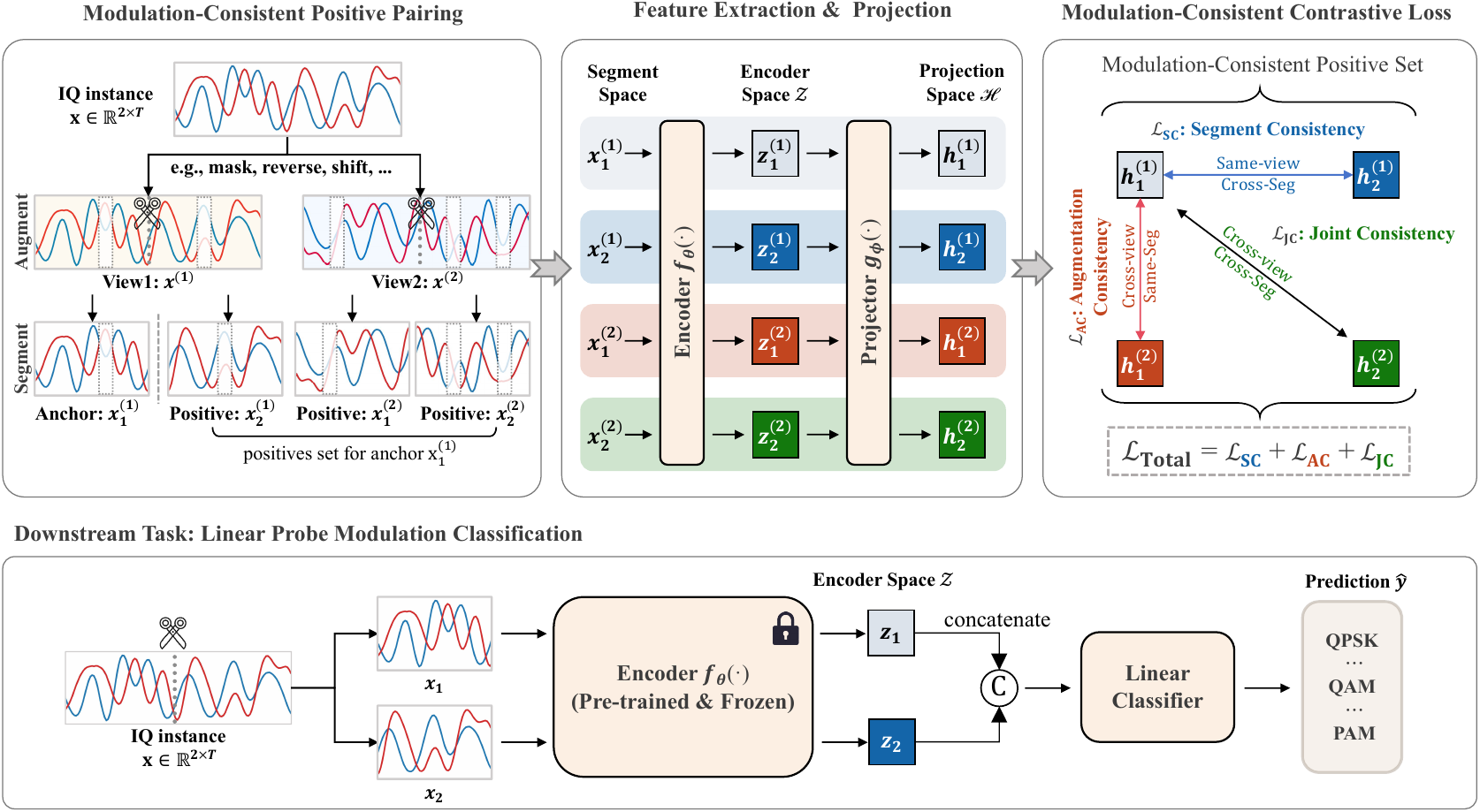}
	\caption{
	The architecture of the proposed Mod-CL framework. The pre-training stage begins with modulation-consistent positive pairing, where an IQ instance $\mathbf{x}$ is transformed into two augmented views $\mathbf{x}^{(1)}$ and $\mathbf{x}^{(2)}$, each subsequently partitioned into two temporal segments $\{ \mathbf{x}_s^{(v)} \}$ that inherently share the same modulation category. 
	These four segments are processed by a shared encoder $f_\theta(\cdot)$ and a shared projector $g_\phi(\cdot)$ to extract encoder features  $\mathbf{z}^{(v)}_{s}\in\mathcal{Z}$ and projected embeddings $\mathbf{h}^{(v)}_{s}\in\mathcal{H}$, respectively. 
	The model is optimized by minimizing a joint loss $\mathcal{L}_{\text{Total}}$ consisting of segment consistency $\mathcal{L}_{\text{SC}}$, augmentation consistency $\mathcal{L}_{\text{AC}}$, and joint consistency $\mathcal{L}_{\text{JC}}$. For downstream evaluation, the pretrained encoder is frozen, and the features of two non-augmented segments are concatenated for linear classification.
	}

	\label{fig:pipeline}
\end{figure*}

\subsection{Overview of the Proposed Framework}
\label{sec:overview}

Building upon the intra-instance modulation consistency prior, this paper develops Mod-CL, a modulation-consistency-based contrastive learning framework to extract modulation-relevant information from unlabeled signals while suppressing symbol-realization-induced interference.
As illustrated in Fig.~\ref{fig:pipeline}, each instance is transformed into multiple segment-level views via data augmentation and temporal partitioning. 
Specifically, positive pairs are formed from different temporal segments of the same received signal, which are expected to share the same modulation type while exhibiting different symbol realizations and local temporal patterns. This label-free positive construction encourages the model to learn modulation-invariant representations.
We further formulate three complementary contrastive objectives based on distinct positive relations induced by segmentation and augmentation, and optimize them in a grouped manner.

In the following subsections, we detail the segment construction and positive pairing strategy (Section~\ref{sec:segment_pairing}), present the contrastive objectives (Section~\ref{sec:objective}), summarize the overall training procedure (Section~\ref{sec:training}), and provide a mechanism-level analysis that supports the proposed design (Section~\ref{sec:mechanism_mi}).

\subsection{Modulation Consistency-Based Contrastive Learning}
\label{sec:segment_pairing}

To fully leverage the intra-instance modulation consistency prior, we develop a modulation-aware positive pairing scheme by contrasting diverse stochastic augmentation policies and symbol sequences.
Given a received IQ instance $\mathbf{x}_n \in \mathbb{R}^{2 \times T}$, we generate a set of four correlated segments $\mathcal{X}_n = \{\mathbf{x}^{(v)}_{n,s} \mid v,s \in \{1,2\}\}$ through a two-stage transformation pipeline:
\subsubsection{Augmentation-induced Variation}
Two stochastic views, $\mathbf{x}_n^{(1)}$ and $\mathbf{x}_n^{(2)}$, are generated via independent signal augmentation functions $a_1(\cdot)$ and $a_2(\cdot)$. These operations (e.g., phase rotation, frequency shifting, and noise injection) simulate common wireless channel impairments while strictly preserving the underlying modulation semantics \cite{xing2025aflnet}.
\subsubsection{Temporal-segmentation Variation}
Each augmented view $\mathbf{x}^{(v)}_{n}$ is further partitioned along the time dimension into two contiguous segments: $\mathbf{x}^{(v)}_{n,1} \triangleq \mathbf{x}^{(v)}_n(:,1\!:\!L)$ and $\mathbf{x}^{(v)}_{n,2} \triangleq \mathbf{x}^{(v)}_n(:,L\!+\!1\!:\!T)$, where $L = \lfloor T/2 \rfloor$. Although these segments exhibit distinct local waveform profiles due to the stochastic nature of symbol realizations, they are fundamentally governed by the same stationary modulation process. This consistent underlying distribution provides a task-aligned self-supervisory cue for learning modulation-specific representations.

The resulting segments are mapped to the latent space via a shared encoder
$f_\theta(\cdot)$ and a projection head $g_\phi(\cdot)$:
\begin{equation}
	\mathbf{z}^{(v)}_{n,s}
	=
	f_\theta\left(\mathbf{x}^{(v)}_{n,s}\right),
	\quad
	\mathbf{h}^{(v)}_{n,s}
	=
	g_\phi\left(\mathbf{z}^{(v)}_{n,s}\right),
\end{equation}
where $\mathbf{z}^{(v)}_{n,s}$ denotes the encoder feature and
$\mathbf{h}^{(v)}_{n,s}$ denotes the projected embedding. The normalized
embedding used for contrastive learning is defined as
\begin{equation}
	\tilde{\mathbf{h}}^{(v)}_{n,s}
	=
	\frac{
		\mathbf{h}^{(v)}_{n,s}
	}{
		\left\|
		\mathbf{h}^{(v)}_{n,s}
		\right\|_2
	}.
\end{equation}
All contrastive losses are computed using
$\tilde{\mathbf{h}}^{(v)}_{n,s}$.

This construction naturally induces three tiers of
\textit{modulation-consistent positive relations} within $\mathcal{X}_n$:
\begin{itemize}
	\item \textit{Segment Consistency}: Defined between different temporal
	segments within the same augmented view,
	$(\tilde{\mathbf{h}}^{(v)}_{n,1},
	\tilde{\mathbf{h}}^{(v)}_{n,2})$, capturing invariance to
	symbol-realization-induced variations.
	
	\item \textit{Augmentation Consistency}: Defined between the same
	temporal segment under different augmented views,
	$(\tilde{\mathbf{h}}^{(1)}_{n,s},
	\tilde{\mathbf{h}}^{(2)}_{n,s})$, promoting robustness to
	augmentation-induced perturbations.
	
	\item \textit{Joint Consistency}: Defined between cross-view
	cross-segment pairs,
	$(\tilde{\mathbf{h}}^{(v)}_{n,s},
	\tilde{\mathbf{h}}^{(v')}_{n,s'})$ with $v\neq v'$ and $s\neq s'$,
	modeling consistency under simultaneous augmentation and segment
	variations.
\end{itemize}

\subsection{Modulation-Consistent Contrastive Objectives}
\label{sec:objective}

Based on the structured relations defined in Sec.~\ref{sec:segment_pairing},
we formulate a modulation-consistent contrastive objective. 
For each signal instance, the proposed framework generates four segment-level views, i.e., two augmented views and two temporal segments. These views share the same modulation semantics and thus form modulation-consistent positives. Although the four views of the same instance are all modulation-consistent, we decompose their relations into three structured single-positive contrastive terms, so that the roles of augmentation, segmentation, and their coupled variation can be explicitly modeled.

To avoid repetitive notation, we define a common candidate set and
normalization term for all three contrastive losses. Let
\begin{equation}
	\mathcal{I}
	=
	\{1,\ldots,B\}\times\{1,2\}\times\{1,2\}
\end{equation}
denote the index set of all segment-level views in a mini-batch. 
Let $\operatorname{sim}(\mathbf{u},\mathbf{v})=\mathbf{u}^{\top}\mathbf{v}/\tau$ denote the temperature-scaled cosine similarity. 
For an anchor $\tilde{\mathbf{h}}^{(v)}_{n,s}$, the NT-Xent normalization term is defined as
\begin{equation}
	\begin{aligned}
		D^{(v)}_{n,s}
		=
		\sum_{(n',v',s')\in
			\mathcal{I}\setminus\{(n,v,s)\}}
		\exp
		\left(
		\operatorname{sim}
		\left(
		\tilde{\mathbf{h}}^{(v)}_{n,s},
		\tilde{\mathbf{h}}^{(v')}_{n',s'}
		\right)
		\right).
	\end{aligned}
\end{equation}
The term $D^{(v)}_{n,s}$ aggregates the similarities between the anchor
and all other views in the mini-batch. It includes the corresponding
positive view and other non-anchor views, while excluding only the anchor itself. The three consistency losses share this normalization term and differ in the choice of modulation-consistent positive pairs.

\paragraph{Augmentation Consistency Loss ($\mathcal{L}_{\mathrm{AC}}$)}
This term aligns the same temporal segment across different augmented
views, encouraging invariance to non-semantic channel-style perturbations.
For anchor $\tilde{\mathbf{h}}^{(1)}_{n,s}$, the positive counterpart is
$\tilde{\mathbf{h}}^{(2)}_{n,s}$. The loss is defined as
\begin{equation}
	\mathcal{L}_{\mathrm{AC}}
	=
	-\frac{1}{2B}
	\sum_{n=1}^{B}
	\sum_{s=1}^{2}
	\log
	\frac{
		\exp
		\left(
		\operatorname{sim}
		\left(
		\tilde{\mathbf{h}}^{(1)}_{n,s},
		\tilde{\mathbf{h}}^{(2)}_{n,s}
		\right)
		\right)
	}{
		D^{(1)}_{n,s}
	}.
\end{equation}

\paragraph{Segment Consistency Loss ($\mathcal{L}_{\mathrm{SC}}$)}
This term aligns different temporal segments within the same augmented
view, encouraging invariance to symbol-realization-induced variations.
For anchor $\tilde{\mathbf{h}}^{(v)}_{n,1}$, the positive counterpart is
$\tilde{\mathbf{h}}^{(v)}_{n,2}$. The loss is defined as
\begin{equation}
	\mathcal{L}_{\mathrm{SC}}
	=
	-\frac{1}{2B}
	\sum_{v=1}^{2}
	\sum_{n=1}^{B}
	\log
	\frac{
		\exp
		\left(
		\operatorname{sim}
		\left(
		\tilde{\mathbf{h}}^{(v)}_{n,1},
		\tilde{\mathbf{h}}^{(v)}_{n,2}
		\right)
		\right)
	}{
		D^{(v)}_{n,1}
	}.
\end{equation}

\paragraph{Joint Consistency Loss ($\mathcal{L}_{\mathrm{JC}}$)}
The above two losses impose consistency along the augmentation and segment
dimensions separately. To further regularize their coupled variation, we
align cross-view cross-segment pairs. Specifically, for each instance,
$\tilde{\mathbf{h}}^{(1)}_{n,1}$ is paired with $\tilde{\mathbf{h}}^{(2)}_{n,2}$, leading to
\begin{equation}
	\mathcal{L}_{\mathrm{JC}}
	=
	-\frac{1}{B}
	\sum_{n=1}^{B}
	\log
	\frac{
		\exp
		\left(
		\operatorname{sim}
		\left(
		\tilde{\mathbf{h}}^{(1)}_{n,1},
		\tilde{\mathbf{h}}^{(2)}_{n,2}
		\right)
		\right)
	}{
		D^{(1)}_{n,1}
	}.
\end{equation}

The final pre-training objective is
\begin{equation}
	\label{eq:l_total}
	\mathcal{L}_{\mathrm{total}}
	=
	\mathcal{L}_{\mathrm{SC}}
	+
	\mathcal{L}_{\mathrm{AC}}
	+
	\mathcal{L}_{\mathrm{JC}}.
\end{equation}
By jointly enforcing consistency across augmentation, segmentation, and
their coupled variation, the proposed objective encourages the encoder to
capture modulation-discriminative structures while suppressing
non-semantic channel perturbations and symbol-level waveform variations.

\subsection{Overall Pretraining and Downstream Evaluation}
\label{sec:training}

We summarize the self-supervised pretraining pipeline of the proposed framework.
In each iteration, we sample a mini-batch of $B$ received signal instances $\{\mathbf{x}_n\}_{n=1}^{B}$.
For each instance, two stochastic augmented views $\mathbf{x}^{(1)}_n$ and $\mathbf{x}^{(2)}_n$ are generated and each view is partitioned into two contiguous temporal segments, yielding four segments per instance.
All segments are then processed by the shared encoder $f_{\theta}(\cdot)$ and projection head $g_{\phi}(\cdot)$ to obtain projected embeddings, which are further $\ell_2$-normalized for contrastive learning.
Based on these normalized embeddings, we compute the three contrastive losses $\mathcal{L}_{\mathrm{SC}}$, $\mathcal{L}_{\mathrm{AC}}$, and $\mathcal{L}_{\mathrm{JC}}$, and optimize their sum in \eqref{eq:l_total} via backpropagation.
The detailed procedure is summarized in Algorithm~\ref{alg:pretrain}.

\begin{algorithm}[t]
	\caption{Self-Supervised Pretraining of Mod-CL}
	\label{alg:pretrain}
	\begin{algorithmic}[1]
		\REQUIRE Unlabeled dataset $\mathcal{D}$; encoder $f_{\theta}$; projection head $g_{\phi}(\cdot)$; temperature $\tau$; $\mathrm{norm}(\cdot)$ denotes $\ell_2$ normalization
		\WHILE{not converged}
		\STATE Sample a mini-batch $\{\mathbf{x}_n\}_{n=1}^{B}$ from $\mathcal{D}$
		\FOR{$n=1$ to $B$}
		\STATE Generate two augmented views:
		\STATE \hspace{0.65em} $\mathbf{x}^{(1)}_n \leftarrow a_1(\mathbf{x}_n), \quad \mathbf{x}^{(2)}_n \leftarrow a_2(\mathbf{x}_n)$
		\STATE Partition each view into two temporal segments:
		\STATE \hspace{0.65em} $\mathbf{x}^{(v)}_{n,1} \leftarrow \mathbf{x}^{(v)}_n(:,1\!:\!L), \quad v \in \{1,2\}$
		\STATE \hspace{0.65em} $\mathbf{x}^{(v)}_{n,2} \leftarrow \mathbf{x}^{(v)}_n(:,L\!+\!1\!:\!T), \quad v \in \{1,2\}$
		
		\STATE Compute encoder features, projected embeddings, and normalized embeddings:
		\STATE $\mathbf{z}^{(v)}_{n,s}
		\leftarrow
		f_\theta\left(\mathbf{x}^{(v)}_{n,s}\right),\quad s\in\{1,2\}$
		\STATE $\mathbf{h}^{(v)}_{n,s}
		\leftarrow
		g_\phi\left(\mathbf{z}^{(v)}_{n,s}\right),\quad s\in\{1,2\}$
		\STATE $\tilde{\mathbf{h}}^{(v)}_{n,s}
		\leftarrow
		\operatorname{norm}
		\left(
		\mathbf{h}^{(v)}_{n,s}
		\right),\quad s\in\{1,2\}$		
		\ENDFOR
		\STATE Compute $\mathcal{L}_{\mathrm{SC}}$, $\mathcal{L}_{\mathrm{AC}}$,
		and $\mathcal{L}_{\mathrm{JC}}$ using
		$\tilde{\mathbf{h}}^{(v)}_{n,s}$
		\STATE $\mathcal{L}_{\mathrm{Total}} \leftarrow \mathcal{L}_{\mathrm{SC}} + \mathcal{L}_{\mathrm{AC}} + \mathcal{L}_{\mathrm{JC}}$
		\STATE Update encoder and projection head parameters by minimizing $\mathcal{L}_{\mathrm{Total}}$
		\ENDWHILE
		\RETURN Pretrained encoder $f_{\theta}$ for downstream linear probing
	\end{algorithmic}
\end{algorithm}

Upon completion of the pre-training phase, the projection head $g_{\phi}(\cdot)$ is removed, and the encoder $f_{\theta}(\cdot)$ is fixed as a generic feature extractor. To evaluate the quality of the learned representations, we adopt the standard linear probing protocol, where a linear classifier $c_{\omega}(\cdot)$ is trained atop the frozen encoder using a limited set of labeled samples. During this downstream stage, data augmentation is disabled to maintain signal integrity. Following the same temporal partitioning as in the pre-training stage, each IQ instance $\mathbf{x}_n$ is divided into two segments, $\{\mathbf{x}_{n,1}, \mathbf{x}_{n,2}\}$. These segments are independently processed by the frozen encoder to yield segment-level embeddings:
\begin{equation}
	\mathbf{z}_{n,s} = f_{\theta}(\mathbf{x}_{n,s}), \quad s \in \{1, 2\}.
\end{equation}
The two segment-level features are concatenated as
\begin{equation}
	\mathbf{z}_{n}
	=
	[\mathbf{z}_{n,1};\mathbf{z}_{n,2}].
\end{equation}

Finally, $\mathbf{z}_n$ is fed into the linear classifier $c_{\omega}(\cdot)$ to predict the modulation type, with the entire classification head optimized via the standard cross-entropy loss.

\subsection{Mechanism Analysis: Why Segment-Level Pairing Improves AMC}
\label{sec:mechanism_mi}

This subsection explains why segment-level positive pairing is more modulation-aligned than instance-level contrastive learning. The key insight is that contrastive learning encourages agreement between positive pairs, so the information they share determines what the encoder preserves.
By examining what is shared across positive pairs in each approach, we can understand the inductive bias induced by our design.

\subsubsection{Factorized Signal Model}
We describe the $s$-th temporal segment of the $n$-th IQ instance using a factorized abstraction:
\begin{equation}
	\mathbf{x}_{n,s} = \mathcal{G}(M_n, B_{n,s}, C_n),
	\label{eq:gen_model_factorized}
\end{equation}
where $M_n$ denotes the modulation semantics, $B_{n,s}$ denotes the local symbol realization within the $s$-th segment, and $C_n$ represents instance-level reception conditions (e.g., channel response, SNR, noise statistics, and hardware distortions). Within a short observation window, $M_n$ remains fixed across segments, and $C_n$ is typically shared or highly correlated. In contrast, $B_{n,s}$ generally varies across segments and induces segment-specific local waveform patterns.
\paragraph{What Is Shared and What Is Not}
For Mod-CL, positive pairs are constructed from different temporal segments of the same instance. Following Eq.~\eqref{eq:gen_model_factorized}, two such segments are expressed as:
\begin{equation}
	\mathbf{x}_{n,1} = \mathcal{G}(M_n, B_{n,1}, C_n), \quad
	\mathbf{x}_{n,2} = \mathcal{G}(M_n, B_{n,2}, C_n).
	\label{eq:two_segment_factorized}
\end{equation}
Here, $M_n$ and $C_n$ are shared across segments, while the symbol realizations $B_{n,1}$ and $B_{n,2}$ are not. 
Hence, the mutual information $I(\mathbf{x}_{n,1}; \mathbf{x}_{n,2})$
mainly arises from the shared factors $M_n$ and $C_n$, rather than from
the segment-specific symbol realizations $\mathbf{B}_{n,1}$ and
$\mathbf{B}_{n,2}$.

For instance-level contrastive learning (e.g., SimCLR), positive pairs are two augmented views derived from the same waveform realization:
\begin{equation}
	\begin{aligned}
		\mathbf{x}^{(1)} &= a_1(\mathcal{G}(M_n, B_n, C_n)), \\
		\mathbf{x}^{(2)} &= a_2(\mathcal{G}(M_n, B_n, C_n)).
	\end{aligned}
\end{equation}
where \(B_n\) denotes the full symbol sequence. Both views share the \textbf{same} symbol realization \(B_n\); therefore, the shared information includes \(B_n\) in addition to \(M_n\) and \(C_n\).

\subsubsection{Information Flow to Encoder Representations}
For encoder representations \(\mathbf{z}_{n,s} = f_{\theta}(\mathbf{x}_{n,s})\), the data processing inequality gives:
\begin{multline}
	I(\mathbf{z}_{n,1}; \mathbf{z}_{n,2})
	\le I(\mathbf{x}_{n,1}; \mathbf{x}_{n,2}) \\
	\le \min\!\bigl\{
	I(\mathbf{x}_{n,1}; M_n, C_n),\,
	I(\mathbf{x}_{n,2}; M_n, C_n)
	\bigr\}.
	\label{eq:mi_shared_bound}
\end{multline}
This bound shows that the mutual information between segment-level positive pairs cannot exceed what is shared by \((M_n, C_n)\). Since \(B_{n,1}\) and \(B_{n,2}\) are not shared across segments, they are naturally excluded from the information that contrastive learning can maximize. For instance-level positive pairs, no such exclusion occurs as \(B_n\) is fully shared.

\subsubsection{Why Segment-Level Pairing Produces Cleaner Representations}
The above analysis reveals the fundamental difference between the two strategies. 
For Mod-CL, the contrastive objective encourages the encoder to emphasize
the shared factors $M_n$ and $C_n$ across segment-level positive pairs,
while reducing its reliance on segment-specific symbol realizations
${B}_{n}$.
 For SimCLR, the encoder is encouraged to preserve \(B_n\) alongside \(M_n\) and \(C_n\), leading to representations entangled with symbol-induced nuisance factors. While both \(M_n\) and \(C_n\) are shared in Mod-CL, the downstream linear classifier can exploit the modulation-relevant information \(M_n\) while ignoring channel conditions \(C_n\) if they are not discriminative for the task. 
 The advantage of segment-level pairing lies in its ability to construct positive pairs that share modulation semantics while differing in segment-specific symbol realizations. This design encourages the encoder to focus on stable modulation-discriminative structures across segments, thereby producing representations that are more aligned with the AMC objective.

\subsubsection{Connection to the Proposed Objective}
The three loss terms in Mod-CL instantiate the above principle. The segment consistency loss $\mathcal{L}_{\mathrm{SC}}$ directly enforces the exclusion of symbol-realization information by aligning different temporal segments—the core insight from the information-theoretic analysis. The augmentation consistency loss $\mathcal{L}_{\mathrm{AC}}$ complements this by promoting invariance to channel distortions $C_n$, ensuring that the shared factors captured by the encoder are not dominated by nuisance channel variations. Finally, the joint consistency loss $\mathcal{L}_{\mathrm{JC}}$ regularizes the compounded case where both segment and augmentation vary, providing additional pressure to discard non-modulation factors. Together, these three terms encourage the encoder to focus on modulation-discriminative features while suppressing symbol-induced interference.

\section{Experiments}
\label{sec:exp}
In this section, we detail the construction of the simulated datasets and systematically evaluate the effectiveness of Mod-CL through a series of experiments.  All experiments are implemented in PyTorch~2.0.1 with Python~3.8 and conducted on a single NVIDIA GeForce RTX~3090 GPU with 24~GB memory.

\begin{table*}[t]
	\centering
	\setlength{\tabcolsep}{5pt}
	\renewcommand{\arraystretch}{1.1}
	\caption{
		Overall test accuracy (\%) under the linear probing protocol on RML 2016.10A and RML 2016.10B with different label budgets.
		For SSL methods, the encoder is first pretrained without labels and then frozen, and only a linear classifier is trained using the labeled subset.
		The Random-Init. baseline follows the same linear probing protocol but does not load pretrained encoder weights.
		Here, $N$ denotes the number of labeled samples per modulation class per SNR level used to train the linear classifier.
		Results are reported as mean $\pm$ standard deviation over five random seeds.
		For each dataset and each label budget, the best result is highlighted in
		bold and the second-best result is underlined.
	}
	\label{tab:overall_acc_N}
	
	\begin{tabular}{l l cccccc}
		\toprule
		\multicolumn{8}{c}{\textbf{RML 2016.10A}} \\
		\midrule
		\textbf{Category} & \textbf{Method} & \textbf{2} & \textbf{5} & \textbf{10} & \textbf{20} & \textbf{50} & \textbf{100} \\
		\midrule
		
		\multirow{1}{*}{Random baseline}
		& Random Init.
		& 13.59$\pm$0.70 & 16.08$\pm$0.56 & 16.95$\pm$0.27 & 18.23$\pm$0.21 & 20.13$\pm$0.60 & 21.52$\pm$0.71 \\
		
		\midrule
		
		\multirow{2}{*}{Generic SSL}
		& SimCLR\cite{chen2020simple}  & \underline{45.52$\pm$1.26} & \underline{47.95$\pm$1.03} & \underline{49.07$\pm$0.90} & \underline{50.37$\pm$0.57} & 51.68$\pm$0.41 & 52.86$\pm$0.36 \\
		& MoCo\cite{he2020momentum}    & 32.91$\pm$2.81 & 38.60$\pm$1.01 & 41.30$\pm$1.62 & 42.83$\pm$1.23 & 44.32$\pm$1.55 & 45.56$\pm$1.67 \\
		
		\midrule
		
		\multirow{4}{*}{AMC-Tailored SSL}
		& SemiAMC\cite{liu2021self}  & 36.32$\pm$2.77 & 43.82$\pm$0.45 & 47.51$\pm$0.42 & 49.91$\pm$0.53 & 53.90$\pm$0.37 & 56.03$\pm$0.11 \\
		& MAC\cite{li2025multi}     & 27.92$\pm$1.03 & 35.40$\pm$0.91 & 41.65$\pm$0.82 & 47.76$\pm$0.75 & \underline{54.53$\pm$0.63} & \underline{58.13$\pm$0.69} \\
		
		& \textbf{Mod-CL (Ours)}
		& \textbf{51.76$\pm$0.93} & \textbf{57.33$\pm$0.48} & \textbf{59.10$\pm$0.43} & \textbf{60.32$\pm$0.24} & \textbf{61.40$\pm$0.22} & \textbf{61.88$\pm$0.25} \\
		
		\bottomrule
	\end{tabular}
	
	\vspace{1.2ex}
	
	\begin{tabular}{l l cccccc}
		\toprule
		\multicolumn{8}{c}{\textbf{RML 2016.10B}} \\
		\midrule
		\textbf{Category} & \textbf{Method} & \textbf{2} & \textbf{5} & \textbf{10} & \textbf{20} & \textbf{50} & \textbf{100} \\
		\midrule
		
		\multirow{1}{*}{Random baseline}
		& Random Init.
		& 13.37$\pm$1.39 & 15.88$\pm$0.23 & 17.18$\pm$0.20 & 18.26$\pm$0.28 & 20.05$\pm$0.47 & 21.51$\pm$0.86 \\
		
		\midrule
		
		\multirow{2}{*}{Generic SSL}
		& SimCLR\cite{chen2020simple}  & \underline{47.03$\pm$1.17} & \underline{49.16$\pm$1.03} & \underline{50.66$\pm$0.39} & \underline{51.89$\pm$0.42} & \underline{53.23$\pm$0.48} & 54.34$\pm$0.59 \\
		& MoCo\cite{he2020momentum}    & 38.07$\pm$1.98 & 42.25$\pm$0.68 & 43.44$\pm$0.47 & 44.13$\pm$0.44 & 45.44$\pm$0.59 & 46.55$\pm$0.55 \\
		
		\midrule
		
		\multirow{4}{*}{AMC-Tailored SSL}
		& SemiAMC\cite{liu2021self} & 31.61$\pm$3.13 & 39.99$\pm$2.61 & 44.14$\pm$1.45 & 48.24$\pm$0.70 & 52.62$\pm$0.06 & 55.52$\pm$0.36 \\	
		& MAC\cite{li2025multi}     & 20.63$\pm$0.93 & 26.92$\pm$0.85 & 32.73$\pm$0.72 & 38.82$\pm$0.72 & 51.00$\pm$0.76 & \underline{57.51$\pm$0.71} \\
		
		& \textbf{Mod-CL (Ours)}
		& \textbf{53.22$\pm$1.43} & \textbf{58.09$\pm$1.03} & \textbf{60.36$\pm$0.35} & \textbf{61.80$\pm$0.25} & \textbf{62.99$\pm$0.21} & \textbf{63.65$\pm$0.19} \\
		
		\bottomrule
	\end{tabular}
\end{table*}

\subsection{Experimental Setup}
\label{sec:exp_setup}

To evaluate the proposed method, we adopt the standard linear probing protocol widely used in self-supervised learning~\cite{oord2018representation}, which is particularly suitable for assessing representation quality under limited supervision. Specifically, the encoder is first pretrained on unlabeled data with the self-supervised objective for 240 epochs, and then frozen while a linear classifier is trained for 80 epochs using labeled data.

Experiments are conducted on two public AMC benchmark datasets, namely RadioML 2016.10A and RadioML 2016.10B~\cite{o2016convolutional}. Following~\cite{li2025multi}, each dataset is randomly split into training, validation, and test sets with a ratio of 6:1:3. All labeled samples used for linear probing are drawn exclusively from the training split to avoid data leakage.
Unless otherwise specified, all methods use the same M2SFE-tiny encoder backbone~\cite{wang2024sigda} to ensure a fair comparison. During self-supervised pretraining, an MLP projection head is attached to the encoder and discarded after pretraining. The projection head consists of two linear layers with a hidden dimension of 256 and an output dimension of 128, with Batch Normalization and LeakyReLU applied between them. The contrastive temperature is set to $\tau = 0.07$.
For Mod-CL, the input signal is evenly partitioned into two segments of length 64 by default, which are used to construct modulation-consistent positive pairs for contrastive learning.

For optimization, we use Adam with an initial learning rate of $1\times10^{-3}$ and a mini-batch size of 256, without additional learning-rate decay or early stopping. For each signal-to-noise ratio (SNR) level, $N$ labeled samples are selected for each modulation class to train the linear classifier, where $N \in \{2, 5, 10, 20, 50, 100\}$.
Unless otherwise specified, all reported accuracies are test accuracies (\%) under the linear probing protocol, averaged over five random seeds, with mean $\pm$ standard deviation reported when applicable. 
For the ablation studies, unless otherwise stated, experiments are conducted on RML 2016.10A under the setting with $N=5$.

To construct diverse yet modulation-preserving views, we apply a set of stochastic signal-level augmentations to raw IQ samples, each activated with probability 0.5. 
Specifically, random masking removes a portion of temporal samples by zeroing short segments; amplitude scaling rescales the signal to simulate power variations; time shifting applies a circular temporal shift; phase rotation rotates the IQ signal in the complex plane; and sign inversion multiplies the signal by $-1$.

\subsection{Comparative Experiment}

\begin{figure*}
	\centering
	\includegraphics[width=0.9\linewidth]{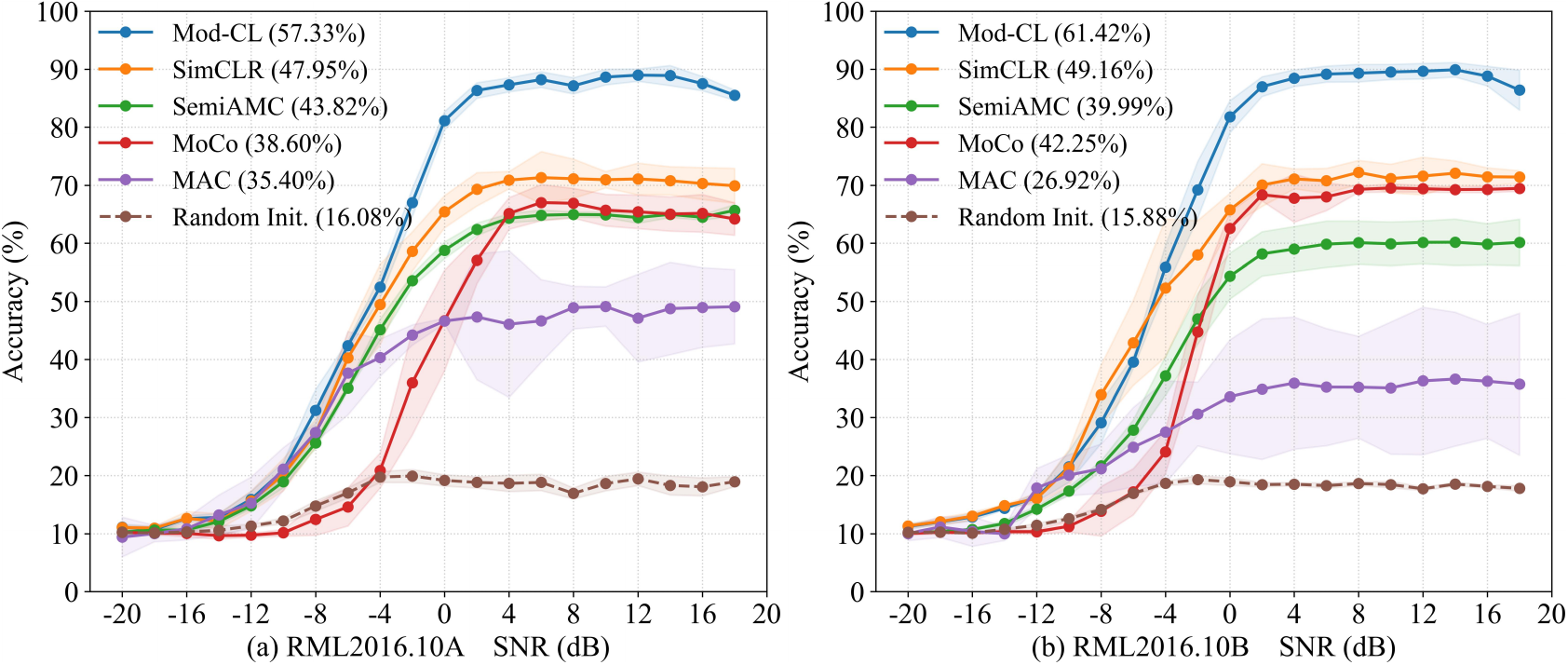}
	\caption{Per-SNR test accuracy (\%) under the linear probing protocol with $N=5$ labeled samples per class per SNR on RML 2016.10A and RML 2016.10B. The encoder is frozen and only a linear classifier is trained. Curves are averaged over five random seeds; shaded regions denote one standard deviation. The numbers in parentheses indicate the overall average test accuracy (\%) under the same label budget.
	}
	\label{fig:comparezhexian}
\end{figure*}

\subsubsection{Overall Comparison Under Linear Probing Protocol}

To assess whether the proposed modulation-aligned pretraining improves downstream linear separability under limited supervision, we compare Mod-CL against representative classical self-supervised methods, including SimCLR~\cite{chen2020simple} and MoCo~\cite{he2020momentum}, as well as AMC-oriented approaches such as MAC~\cite{li2025multi}, and SemiAMC~\cite{liu2021self}, under the same linear probing protocol.
All methods are re-implemented under a unified experimental protocol with identical data splits, label budgets, encoder backbone, and downstream evaluation settings.
For each baseline, we retain its original pretext objective and method-specific design, while keeping common training settings consistent whenever applicable.
In the linear probing experiments, the encoder is frozen and only a single linear classifier is trained to evaluate the discriminative quality of the learned representations.

Table~\ref{tab:overall_acc_N} summarizes the linear probing performance of different self-supervised methods on the RadioML datasets under varying label budgets, where $N$ denotes the number of labeled samples per modulation class per SNR level. 
We additionally conduct paired $t$-tests over five matched random seeds between Mod-CL and the strongest competing baseline for each dataset and each label budget, and all improvements are statistically significant at the $p<0.05$ level.
Across both datasets, Mod-CL consistently achieves the best performance for all values of $N$, showing that the proposed modulation-consistent pretraining strategy learns representations that are more suitable for downstream AMC.

The advantage of Mod-CL is particularly evident in low-label regimes. On RML 2016.10A, Mod-CL achieves $51.76\%$ and $57.33\%$ accuracy with only $N=2$ and $N=5$ labeled samples per class per SNR, respectively, clearly outperforming the strongest competing baseline. A similar trend can be observed on RML 2016.10B, where Mod-CL reaches $53.22\%$ at $N=2$ and $58.09\%$ at $N=5$. These results suggest that intra-instance modulation consistency provides an effective self-supervisory signal, allowing the learned representations to become linearly separable with very limited labeled data.

Compared with generic contrastive methods such as SimCLR and MoCo, Mod-CL shows consistent gains across all label budgets. This indicates that instance-level augmentation-based objectives are less effective at isolating modulation-relevant information from nuisance factors such as symbol realizations, noise, and channel-induced variations. Compared with AMC-oriented SSL methods such as SemiAMC and MAC, Mod-CL also maintains a clear margin, especially when labeled samples are scarce. Although MAC becomes competitive with larger label budgets, e.g., $58.13\%$ at $N=100$ on RML 2016.10A and $57.51\%$ on RML 2016.10B, it remains inferior to Mod-CL. Overall, these results demonstrate that explicitly constructing modulation-consistent positive pairs leads to more discriminative and task-aligned representations for AMC.

\subsubsection{Performance Analysis Under Varying SNR Conditions}
To examine whether the gains of Mod-CL are consistent across noise conditions rather than concentrated at a few favorable SNR levels, we report per-SNR linear probing accuracy with $N=5$ labeled samples per class.
As shown in Fig.~\ref{fig:comparezhexian}, the proposed Mod-CL achieves the best overall trend and clear advantages in most SNR regions, especially in the low-to-medium SNR range.
For example, in the SNR range from approximately $-8$~dB to $0$~dB, Mod-CL exhibits a substantially larger margin over both classical contrastive learning methods and AMC-oriented self-supervised baselines.
This behavior suggests that the representations learned by Mod-CL remain more discriminative under severe noise conditions, indicating more effective capture of modulation-discriminative structure during pretraining.

Compared with Mod-CL, baseline methods exhibit a slower improvement as SNR increases.
Classical contrastive learning approaches such as SimCLR and MoCo rely on task-agnostic instance-level objectives, which may be more sensitive to nuisance factors introduced by noise and channel effects.
AMC-oriented self-supervised methods, although incorporating task-specific designs, still exhibit limited robustness in low-SNR conditions, suggesting that their pretext objectives may not explicitly enforce modulation-level invariance across noisy observations.

The consistent superiority of Mod-CL across all SNR levels, together with its pronounced advantage in low- and medium-SNR regimes, provides strong empirical evidence that explicitly leveraging intra-instance modulation consistency leads to noise-resilient and modulation-invariant representations.
These results further validate the effectiveness of grounding self-supervised learning objectives in intra-instance modulation consistency rather than relying solely on generic augmentation-driven objectives.

\subsubsection{Overall Comparison Under Fine-Tuning Protocol}
\label{sec:finetune}

\begin{table}[t]
	\centering
	\setlength{\tabcolsep}{3pt}
	\renewcommand{\arraystretch}{1.08}
	\caption{
		Fine-tuning accuracy (\%) under representative label budgets.
		SSL methods are pretrained for 240 epochs and then fine-tuned end-to-end.
		``Scratch'': end-to-end fine-tuning from random initialization (no pretraining).
		``Fully Sup.'': supervised training with all labeled samples.
		Results are mean $\pm$ std over five random seeds.
	}
	\label{tab:finetune_acc_N}
	\begin{tabular}{l l ccc}
		\toprule
		\textbf{Dataset} & \textbf{Method}
		& \textbf{$N=2$} & \textbf{$N=20$} & \textbf{$N=100$} \\
		\midrule
		
		\multirow{4}{*}{2016.10A}
		& Scratch + FT
		& 35.75$\pm$3.28 & 54.39$\pm$0.48 & 59.42$\pm$0.33 \\
		& SimCLR + FT
		& 49.25$\pm$0.30 & 55.30$\pm$0.49 & 60.15$\pm$0.46 \\
		& \textbf{Mod-CL + FT}
		& \textbf{55.91$\pm$1.15} & \textbf{59.93$\pm$0.27} & \textbf{61.76$\pm$0.56} \\
		\cmidrule(lr){2-5}
		& Fully Sup.
		& \multicolumn{3}{c}{63.07$\pm$0.14} \\
		
		\midrule
		
		\multirow{4}{*}{2016.10B}
		& Scratch + FT
		& 35.29$\pm$2.51 & 53.87$\pm$0.28 & 59.31$\pm$0.17 \\
		& SimCLR + FT
		& 49.82$\pm$0.80 & 54.08$\pm$0.79 & 60.41$\pm$0.53 \\	
		& \textbf{Mod-CL + FT}
		& \textbf{56.62$\pm$1.15} & \textbf{61.71$\pm$0.57} & \textbf{63.57$\pm$0.49} \\
		\cmidrule(lr){2-5}
		& Fully Sup.
		& \multicolumn{3}{c}{64.14$\pm$0.22} \\
		
		\bottomrule
	\end{tabular}
\end{table}

While linear probing evaluates the quality of frozen representations, fine-tuning allows the encoder to adapt to the downstream task through end-to-end optimization. To assess the transferability of the learned features under practical scenarios where limited labeled data are available, we evaluate Mod-CL under the fine-tuning protocol on both RML 2016.10A and RML 2016.10B datasets. Following the same setting as linear probing, we pretrain the encoder without labels and then fine-tune it jointly with a linear classifier using limited labeled samples. We report results under three representative label budgets: $N=2$, $N=20$, and $N=100$ labeled samples per modulation class per SNR level. For comparison, we include a scratch baseline (random initialization, then supervised fine-tuning with the same labeled data) and a fully supervised upper bound (end-to-end training using all labeled training samples, corresponding to $N=600$ per class). 
Among the self-supervised baselines, we select SimCLR for fine-tuning comparison because it is a strong and widely used generic contrastive baseline and shows stable performance in the linear probing experiments.
Table~\ref{tab:finetune_acc_N} summarizes the fine-tuning accuracy of Mod-CL and SimCLR, together with the scratch and fully supervised baselines.

Several observations can be drawn from the results. First, both self-supervised methods significantly outperform the scratch baseline across all label budgets and datasets. For example, on RML 2016.10A with $N=2$, SimCLR achieves $49.25\%$ accuracy compared to scratch's $35.75\%$, demonstrating that self-supervised pretraining provides a strong initialization for downstream adaptation. Second, Mod-CL consistently outperforms SimCLR in all settings. The advantage is most pronounced under extremely limited supervision: on RML 2016.10A with $N=2$, Mod-CL achieves $55.91\%$ accuracy, outperforming SimCLR ($49.25\%$) by $6.66\%$. On RML 2016.10B, the gain is even larger ($56.62\%$ vs. $49.82\%$, a $6.80\%$ improvement). These results mirror the trend observed in linear probing and further confirm that the modulation-aligned pretraining objective yields more transferable representations.

Third, as the label budget increases, the performance gap between Mod-CL and SimCLR gradually narrows. On RML 2016.10A, the gap decreases from $6.66\%$ at $N=2$ to $4.63\%$ at $N=20$ and $1.61\%$ at $N=100$. This is expected, as abundant labeled data reduces the reliance on pretrained features, making the advantage of a better initialization less critical. Notably, with only $N=100$ labeled samples, Mod-CL achieves $61.76\%$ accuracy on RML 2016.10A and $63.57\%$ on RML 2016.10B, approaching the fully supervised upper bounds ($63.07\%$ and $64.14\%$, respectively) with a gap of only $1.31\%$ and $0.57\%$. This indicates that Mod-CL effectively captures modulation-discriminative information during pretraining, requiring very few labeled samples to adapt to near-optimal performance.

Overall, the fine-tuning results corroborate the linear probing findings: Mod-CL learns transferable and modulation-aligned representations that excel under limited supervision, achieving competitive performance with minimal labeled data and approaching fully supervised upper bounds with only $N=100$ samples.

\subsection{Representation Analysis}
\label{sec:tsne}

\begin{figure}
	\centering
	\includegraphics[width=0.9\linewidth]{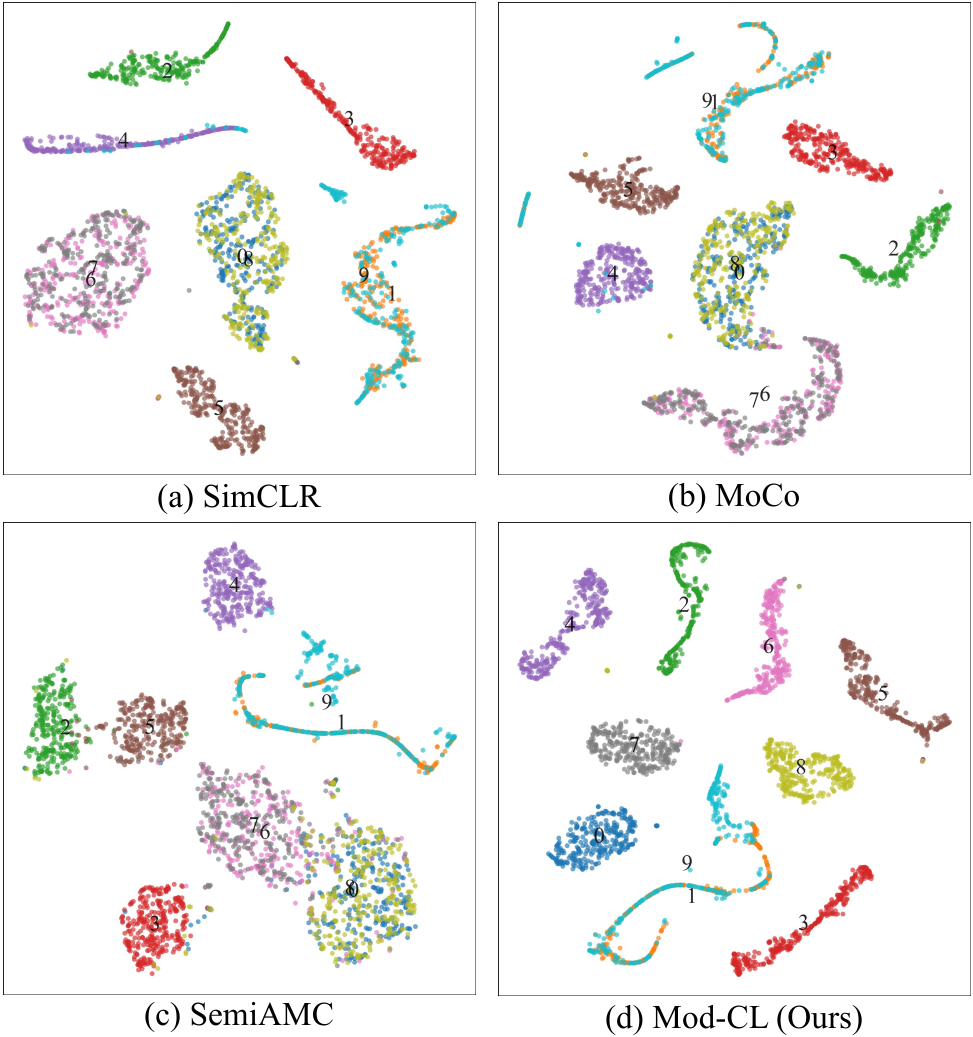}
	\caption{
		t-SNE visualization of encoder features on the test set of RML 2016.10A under the $N=5$ labeled samples per class setting. (a) SimCLR, (b) MoCo, (c) SemiAMC, (d) Mod-CL (ours). All subplots use the same evaluation split and t-SNE configuration with identical random seed.
	}
	\label{fig:tsne}
\end{figure}

To qualitatively assess whether Mod-CL improves class separability in the learned feature space, we visualize encoder representations using t-SNE on the same evaluation split under the $N=5$ labeled samples per class setting. The features are extracted from the encoder outputs and correspond to the representations used by the linear classifier. We compare the proposed Mod-CL method with several representative self-supervised baselines (SimCLR, MoCo, and SemiAMC), using identical t-SNE settings for all methods.

As shown in Fig.~\ref{fig:tsne}, baseline methods exhibit partial clustering behavior; however, noticeable overlaps and ambiguities remain among several modulation classes, reflecting insufficient inter-class separation or poor intra-class compactness. SimCLR (Fig.~\ref{fig:tsne}(a)) and MoCo (Fig.~\ref{fig:tsne}(b)) demonstrate similar trends, where multiple modulation types remain entangled. SemiAMC (Fig.~\ref{fig:tsne}(c)) shows some cluster separation but with notable inter-class mixing. 

In contrast, the proposed Mod-CL method (Fig.~\ref{fig:tsne}(d)) produces significantly more structured embeddings. 
Most modulation types exhibit more compact and better separated visual clusters, while the main overlap appears between AM-DSB and WBFM, which is a known characteristic of the RML 2016.10A dataset, where WBFM signals during silent intervals exhibit patterns similar to AM-DSB, making them inherently difficult to distinguish. Apart from this intrinsic ambiguity, Mod-CL effectively reduces class confusion and yields improved cluster separability compared to all baseline methods.

These qualitative results are consistent with the quantitative improvements reported in the previous sections and further support that Mod-CL learns modulation-discriminative representations that are highly amenable to linear separation under limited labeled supervision.

\subsection{Mechanism Validation}
\label{sec:mechanism_validation}

\begin{figure}[t]
	
	\centering
	\begin{subfigure}{0.23\textwidth}  
		\centering
		\includegraphics[width=\linewidth]{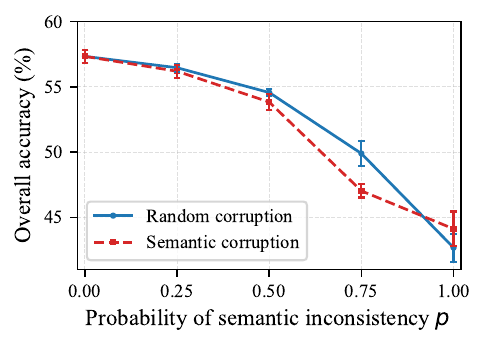}
		\caption{Semantic corruption analysis}
		\label{fig:semantic_corruption}
	\end{subfigure}
	\hfill 
	\begin{subfigure}{0.233\textwidth}
		\centering
		\includegraphics[width=\linewidth]{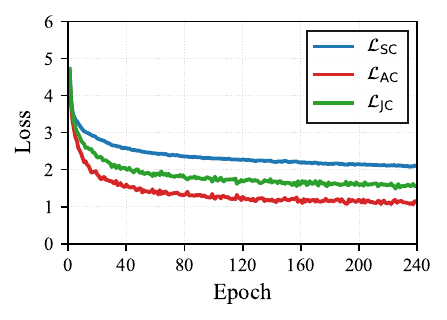}
		\caption{Training curves}
		\label{fig:loss_curves}
	\end{subfigure}
	\caption{Mechanism validation on RML 2016.10A under the linear probing protocol with $N=5$. (a) Semantic corruption analysis. During pretraining, each positive-pair relation is replaced with probability $p$. Curves show overall test accuracy (\%) averaged over five random seeds; error bars indicate $\pm$ one standard deviation. (b) Training curves of $\mathcal{L}_{\mathrm{SC}}$, $\mathcal{L}_{\mathrm{AC}}$, and $\mathcal{L}_{\mathrm{JC}}$ during self-supervised pretraining. All losses decrease steadily and converge within 240 epochs.}
	\label{fig:machanism}
\end{figure}

\paragraph{Causal Effect of Modulation Consistency}
To directly test whether the effectiveness of Mod-CL arises from modulation-consistent positive-pair construction, we perturb the positive-pair sampling process while keeping the architecture, loss definition, and training protocol unchanged. 
This design enables a direct examination of the causal effect of semantic consistency on representation learning. 
Specifically, during pretraining, the original positive pairs, which are constructed from two temporal segments of the same received signal instance, are replaced with probability $p$. By gradually increasing $p \in \{0, 0.25, 0.5, 0.75, 1.0\}$, we systematically evaluate how violations of modulation-level semantic consistency affect downstream modulation classification performance.
Modulation labels are used only to construct the corrupted pairs for this diagnostic analysis and are not used in the proposed self-supervised pretraining.

We consider two corruption strategies:
\begin{itemize}
	\item \textbf{Random corruption}: the paired segment is randomly sampled from the same mini-batch without enforcing modulation constraints;
	\item \textbf{Semantic corruption}: the paired segments are forcibly sampled from different modulation classes, ensuring that their modulation types are entirely mismatched.
\end{itemize}

As shown in Fig.~\ref{fig:machanism}(a), as $p$ increases, downstream classification accuracy declines under both corruption strategies. Notably, semantic corruption causes consistently more severe degradation than random corruption. This is expected, as semantic corruption explicitly enforces wrong modulation semantics, directly contradicting the AMC objective, whereas random corruption may accidentally preserve modulation consistency with some probability. 
These observations provide direct empirical evidence that preserving modulation consistency in positive-pair construction is critical to the effectiveness of Mod-CL.

When positive pairs explicitly enforce conflicting modulation information, the contrastive objective becomes misaligned with the downstream task, resulting in substantial loss of discriminative capability. Even under full corruption ($p=1.0$), the model does not collapse catastrophically, implying that contrastive learning can still capture generic signal characteristics. However, the distinctive advantage of Mod-CL disappears in this case, further confirming that its effectiveness stems from modulation-consistent positive-pair construction rather than architectural or optimization artifacts.

\paragraph{Optimization Stability}
Having established the importance of modulation consistency, we next examine whether the proposed three-term loss is well-optimized in practice. Fig.~\ref{fig:machanism}(b) shows the training curves of $\mathcal{L}_{\mathrm{SC}}$, $\mathcal{L}_{\mathrm{AC}}$, and $\mathcal{L}_{\mathrm{JC}}$ during self-supervised pretraining. All three losses decrease steadily and converge within 240 epochs without noticeable oscillation. The similar decreasing trends of $\mathcal{L}_{\mathrm{SC}}$ and $\mathcal{L}_{\mathrm{JC}}$ suggest that the joint consistency loss does not conflict with the segment consistency loss, but rather reinforces it by regularizing compounded variations. $\mathcal{L}_{\mathrm{AC}}$ converges slightly faster, which is expected as augmentation-based invariance is relatively easier to learn than symbol-realization invariance. 
These observations suggest that the proposed structured loss is stably optimized without evident component conflicts.

\subsection{Ablation Studies}
\label{sec:ablation}
This section presents ablation studies to systematically validate the key design principles of Mod-CL from a design-level perspective. Building upon the preceding mechanism analysis and semantic corruption experiments, which verify the importance of intra-instance modulation consistency, we further decompose the proposed framework to examine how each component contributes to the overall performance.

Specifically, we first analyze the contribution of individual loss components, each encoding a distinct form of segment-level semantic consistency. We then investigate the effect of segmentation granularity, which governs whether the constructed segment pairs retain sufficient modulation-discriminative information. Furthermore, we conduct a label-efficiency analysis to evaluate how effectively Mod-CL leverages limited supervision. Finally, sensitivity analyses with respect to segment length, temperature, batch size, and learning rate are provided to assess training stability and robustness.

\begin{table}[t]
	\centering
	\caption{Ablation Study of Loss Components on the RML 2016.10A Dataset. The model is evaluated under the linear probing protocol with a label budget of $N=5$ per class per SNR level. Results are reported as mean $\pm$ standard deviation over five independent random seeds.}
	\label{tab:ablation_final}
	\setlength{\tabcolsep}{8pt} 
	\renewcommand{\arraystretch}{1.2} 
	\begin{tabular}{cccccc} 
		\toprule
		\textbf{Index} & $\mathcal{L}_{\text{AC}}$ & $\mathcal{L}_{\text{SC}}$ & $\mathcal{L}_{\text{JC}}$ & \textbf{Accuracy (\%)} & \textbf{Gain (\%)} \\
		\midrule
		1 & \checkmark &            &            & 53.73 $\pm$ 0.92 & -- \\
		2 &            & \checkmark &            & 53.72 $\pm$ 1.02 & $-$0.01 \\
		3 &            &            & \checkmark & 55.94 $\pm$ 0.85 & $+$2.21 \\
		\midrule
		4 & \checkmark & \checkmark &            & 56.45 $\pm$ 0.68 & $+$2.72 \\
		5 & \checkmark &            & \checkmark & 56.68 $\pm$ 0.72 & $+$2.95 \\
		6 &            & \checkmark & \checkmark & 56.92 $\pm$ 0.61 & $+$3.19 \\
		\midrule
		\textbf{7} & \checkmark & \checkmark & \checkmark & \textbf{57.33 $\pm$ 0.48} & \textbf{+3.60} \\
		\bottomrule
		\multicolumn{6}{l}{\textsuperscript{*} Index 1 serves as the baseline model for gain calculation.}
	\end{tabular}
\end{table}

\subsubsection{Ablation on Loss Components}
\label{sec:ablation_loss}

Table~\ref{tab:ablation_final} evaluates the contribution of different consistency constraints by progressively incorporating the proposed loss components. 
Using only $\mathcal{L}_{\mathrm{JC}}$ (Index 3) yields 55.94\% accuracy, which already surpasses all compared baselines under the same setting (e.g., SimCLR at 47.95\% in Table~\ref{tab:overall_acc_N}), indicating that the joint consistency objective alone provides a strong supervisory signal. In contrast, using only $\mathcal{L}_{\mathrm{AC}}$ or $\mathcal{L}_{\mathrm{SC}}$ (Indices 1 and 2) achieves lower accuracy (53.73\% and 53.72\%, respectively), suggesting that these two terms alone are less effective but play important supporting roles.
Adding $\mathcal{L}_{\mathrm{SC}}$ to $\mathcal{L}_{\mathrm{AC}}$ (Index 4: 56.45\%) improves the accuracy by 2.72\% over the baseline, demonstrating the complementarity between segment-level and augmentation-level invariance. Combining $\mathcal{L}_{\mathrm{JC}}$ with $\mathcal{L}_{\mathrm{AC}}$ (Index 5: 56.68\%) or with $\mathcal{L}_{\mathrm{SC}}$ (Index 6: 56.92\%) yields further improvements. Finally, the full model incorporating all three losses (Index 7) achieves the best accuracy of 57.33\%, with a gain of 3.60\% over the baseline.

These gains arise because the three loss components capture intra-instance modulation consistency through complementary relational structures across different dimensions. $\mathcal{L}_{\mathrm{SC}}$ enforces consistency across different temporal segments within the same augmented view, encouraging the model to capture modulation semantics that are invariant to symbol-realization-induced variations. $\mathcal{L}_{\mathrm{AC}}$ introduces cross-view consistency for the same temporal segment, promoting robustness to channel-style distortions. $\mathcal{L}_{\mathrm{JC}}$ extends this formulation by modeling cross-view cross-segment relations simultaneously, capturing consistency under compounded variations. Together, these results indicate that jointly modeling consistency across augmentation and segmentation leads to more modulation-discriminative representations.

\subsubsection{Label Efficiency Analysis}
\label{sec:ablation_label_budget}

To quantify how much labeled supervision is needed for the learned representations to become linearly separable, we analyze the label efficiency of Mod-CL across different label budgets on RML 2016.10A. Table~\ref{tab:label_budget_analysis} reports the overall accuracy of Mod-CL and the corresponding gap to the second-best baseline under each label budget.

Mod-CL shows an early saturation trend as the label budget increases. Its accuracy rises rapidly from $51.76\%$ at $N=2$ to $59.10\%$ at $N=10$, while further increasing $N$ to $100$ brings only a modest gain to $61.88\%$. This suggests that the learned representations already capture discriminative modulation information with very limited labels.
The gap to the second-best baseline follows a similar trend. Mod-CL achieves the largest advantage of $10.03$ percentage points at $N=10$, and the gap gradually decreases to $3.75$ points at $N=100$ as more labels become available. These results indicate that Mod-CL provides stronger linear separability in low-label regimes, whereas competing methods require more supervision to approach comparable performance.

These advantages can be attributed to the modulation-aligned pretraining objective, which encourages the encoder to focus on modulation-relevant features while suppressing nuisance variations. As a result, the learned representations are more structured and closer to linearly separable in the embedding space, reducing the dependence on labeled supervision during downstream training.

\begin{table}[t]
	\centering
	\caption{Label-efficiency analysis under the linear probing protocol on RML 2016.10A. Here, $N$ denotes the number of labeled samples per modulation class per SNR level. Results are reported as overall test accuracy (\%) averaged over five random seeds. The gap is computed between Mod-CL and the second-best baseline under each label budget.}
	\label{tab:label_budget_analysis}
	\setlength{\tabcolsep}{5.5pt}
	\begin{tabular}{c|c|c|c}
		\toprule
		\textbf{Label Budget ($N$)} 
		& \textbf{Mod-CL (\%)} 
		& \textbf{Second-best (\%)} 
		& \textbf{Gap (\%)} \\
		\midrule
		2   & 51.76$\pm$0.93 & 45.52$\pm$1.26 & 6.24$\pm$1.03 \\
		5   & 57.33$\pm$0.48 & 47.95$\pm$1.03 & 9.38$\pm$0.83 \\
		10  & 59.10$\pm$0.43 & 49.07$\pm$0.90 & 10.03$\pm$0.69 \\
		20  & 60.32$\pm$0.24 & 50.37$\pm$0.57 & 9.95$\pm$0.51 \\
		50  & 61.40$\pm$0.22 & 54.53$\pm$0.63 & 6.87$\pm$0.42 \\
		100 & 61.88$\pm$0.25 & 58.13$\pm$0.69 & 3.75$\pm$0.83 \\
		\bottomrule
	\end{tabular}
\end{table}

\begin{figure}
	\centering
	\includegraphics[width=0.95\linewidth]{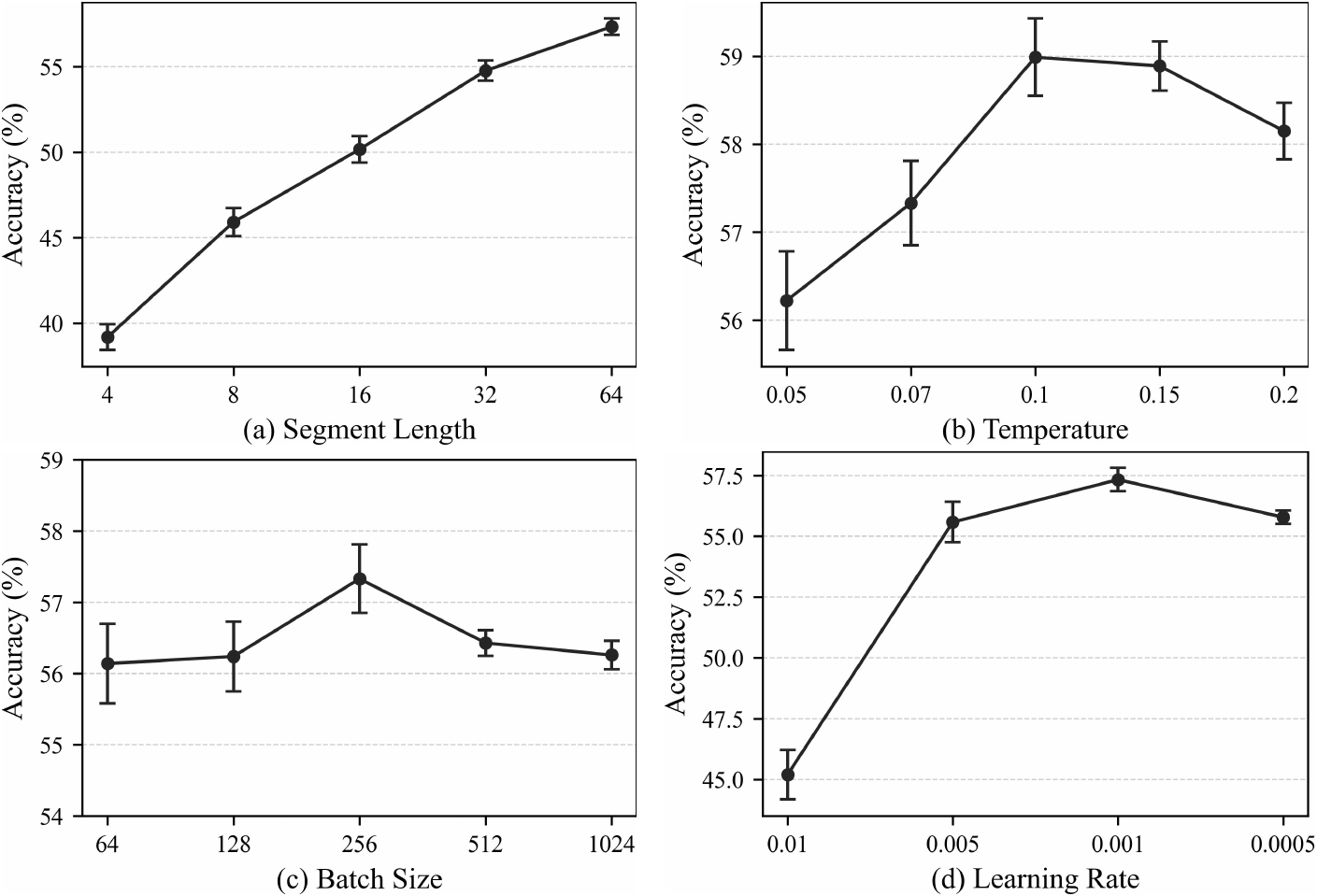}
	\caption{Ablation studies on RML 2016.10A under the linear probing protocol with $N=5$. Results show overall test accuracy (\%) averaged over five random seeds; error bars indicate $\pm$ one standard deviation. (a) Effect of segment length. (b) Effect of temperature. (c) Effect of mini-batch size. (d) Effect of learning rate.}
	\label{fig:ablation}
\end{figure}

\subsubsection{Sensitivity to Hyperparameters}
\label{sec:ablation_hyper}

To evaluate the robustness of Mod-CL to various hyperparameter choices, we systematically investigate the impact of segmentation granularity, temperature, mini-batch size, and learning rate. All results are reported on RML 2016.10A under the linear probing protocol with $N=5$. Fig.~\ref{fig:ablation} summarizes the sensitivity analysis of these four hyperparameters.

\paragraph{Segment Length (Fig.~\ref{fig:ablation}(a))}
We investigate how segmentation length affects downstream accuracy.
As shown in Fig.~\ref{fig:ablation}(a), performance peaks at a segment length of 64 (i.e., two segments), achieving $57.33\%$ accuracy.
Shorter segments cause progressive degradation: accuracy drops to $54.76\%$ at length~32, $50.16\%$ at length~16, $45.91\%$ at length~8, and $39.18\%$ at length~4.
This trend reflects a basic requirement for contrastive learning on modulated signals: each segment must be sufficiently long to retain modulation-discriminative temporal structure.
Excessively short segments lose the semantic information necessary for distinguishing modulation types.

\paragraph{Temperature (Fig.~\ref{fig:ablation}(b))}
We examine the sensitivity of Mod-CL to the temperature parameter $\tau$. As shown in Fig.~\ref{fig:ablation}(b), the accuracy varies modestly across the tested range, increasing from $56.57\%$ at $\tau=0.05$ to $58.99\%$ at $\tau=0.1$, and then slightly decreasing to $58.15\%$ at $\tau=0.2$. This indicates that Mod-CL is robust to temperature variations. Although $\tau=0.1$ gives the best result in this study, we use $\tau=0.07$ in the main comparison experiments for consistency with the common setting adopted by contrastive learning baselines. Under this default setting, Mod-CL still substantially outperforms SimCLR ($47.95\%$), as reported in Table~\ref{tab:overall_acc_N}, suggesting that its advantage mainly stems from the intra-instance modulation consistency prior rather than careful tuning of $\tau$.

\paragraph{Batch Size (Fig.~\ref{fig:ablation}(c))}
We vary the mini-batch size from 64 to 1024. As shown in Fig.~\ref{fig:ablation}(c), batch sizes of 128 and 256 achieve comparable performance, with 256 yielding the highest accuracy ($57.33\%$). Increasing the batch size to 512 or 1024 leads to a slight performance degradation (around $56.4\%$). Notably, despite the variations in batch size, Mod-CL consistently and substantially outperforms the SimCLR baseline ($47.95\%$), suggesting that its performance gains are not sensitive to batch-size tuning.

\paragraph{Learning Rate (Fig.~\ref{fig:ablation}(d))}
We examine the sensitivity of Mod-CL to the learning rate.
As shown in Fig.~\ref{fig:ablation}(d), accuracy peaks at $57.33\%$ with a learning rate of $0.001$, and remains above $55.5\%$ when the rate is halved ($0.0005$) or increased fivefold ($0.005$).
A tenfold increase to $0.01$ causes a marked drop to $45.20\%$, indicating that excessively large steps disrupt pretraining.
Overall, these results demonstrate that Mod-CL is robust to a wide range of hyperparameter choices, and its performance gains are primarily attributed to the proposed modulation-aligned design rather than careful hyperparameter tuning.

\subsubsection{Complexity Analysis}
\label{sec:ablation_complexity}

We further evaluate the parameter count and computational efficiency of Mod-CL to verify that its performance gains do not come from larger model capacity or heavier computation. As summarized in Table~\ref{tab:complexity}, Mod-CL builds upon the SimCLR framework without introducing auxiliary modules, resulting in nearly identical parameter count (0.359M vs.\ 0.360M), GPU memory usage (0.462\,GB vs.\ 0.461\,GB), and training time (12.05\,s vs.\ 12.70\,s per epoch). Under this comparable computational budget, Mod-CL achieves 61.88\% linear probing accuracy, outperforming SimCLR by 10.98 percentage points. This suggests that the gains mainly stem from the proposed intra-instance modulation consistency prior rather than increased model capacity or training overhead. Other methods in Table~\ref{tab:complexity} use comparable or more parameters but achieve lower accuracy, further supporting this conclusion.

\begin{table}[t]
	\centering
	\scriptsize
	\setlength{\tabcolsep}{3pt}
	\renewcommand{\arraystretch}{1.08}
	\caption{
		Complexity comparison of different pretraining methods on RML 2016.10A.
		Pretraining parameters include all modules; inference parameters include only the encoder and classifier.
		Training time and GPU memory are measured with batch size 256.
		Accuracy is linear probing at $N=100$.
	}
	\label{tab:complexity}
	\begin{tabular}{lccccc}
		\toprule
		\textbf{Method}
		& \textbf{Pre. Params}
		& \textbf{Inf. Params}
		& \textbf{Time/Epoch}
		& \textbf{Memory}
		& \textbf{Acc.} \\
		& \textbf{(M)}
		& \textbf{(M)}
		& \textbf{(s)}
		& \textbf{(GB)}
		& \textbf{($N=100$)} \\
		\midrule
		SimCLR  & 0.359 & 0.360 & 12.70 & 0.461 & 52.86 \\
		MoCo    & 0.718 & 0.360 & 11.16 & 0.810 & 45.56 \\
		SemiAMC & 0.373 & 0.391 & 11.92 & 0.365 & 56.03 \\
		\textbf{Mod-CL} & 0.359 & 0.362 & 12.05 & 0.462 & \textbf{61.88} \\
		\bottomrule
	\end{tabular}
\end{table}

\section{Conclusion}

This paper investigates the task-mismatch issue in SSL for AMC, where generic pretext objectives often capture modulation-relevant cues together with nuisance variations, such as symbol realizations and channel noise. To address this issue, we identify intra-instance modulation consistency as a task-aware structural prior: different temporal segments within the same signal instance may exhibit distinct local waveforms but share identical modulation semantics. Building on this prior, we propose Mod-CL, a contrastive framework that encourages the model to capture shared modulation information while suppressing symbol-level interference. Experiments on RadioML datasets demonstrate that Mod-CL achieves competitive performance across various label budgets, with the most pronounced gains in low-label regimes. Semantic corruption tests and ablation studies further validate the effectiveness of modulation-consistent positive pairing. Extending the framework to more diverse datasets and complex real-world channel impairments remains an important direction for future work.


	\bibliographystyle{IEEEtran} 
	\bibliography{reference}

\end{document}